\documentclass[conference]{IEEEtran}
\usepackage{cite}
\usepackage{amsmath,amssymb,amsfonts}
\usepackage{algorithmic}
\usepackage{graphicx}
\usepackage{textcomp}
\usepackage{xcolor}
\def\BibTeX{{\rm B\kern-.05em{\sc i\kern-.025em b}\kern-.08em
    T\kern-.1667em\lower.7ex\hbox{E}\kern-.125emX}}

\usepackage{orcidlink}
\usepackage{svg}
\usepackage{circledsteps}
\usepackage{enumitem}
\usepackage{hyphenat}
\usepackage{csquotes}
\newcommand{\thematicbreak}{\par\bigskip}

\newcommand\helcircled[2][]{\ifmmode
\Circled[fill color=black,inner color=white,#1]{\mathsf{#2}}
\else
\Circled[fill color=black,inner color=white,#1]{\sffamily#2}
\fi
}

\renewcommand{\mkbegdispquote}[2]{\itshape\openautoquote}

\definecolor{darkred}{HTML}{ac162c}
\newcommand{\rev}{}

\begin{document}

\title{Interactive Diagrams for Software Documentation}

\author{%
    \IEEEauthorblockN{%
        Adam Štěpánek\IEEEauthorrefmark{1}\orcidlink{0009-0008-9388-2546},\,%
        David Kuťák\IEEEauthorrefmark{1},\orcidlink{0000-0002-4346-6850}\,%
        Barbora Kozlíková\IEEEauthorrefmark{1}\orcidlink{0000-0003-0045-0872},\,%
        Jan Byška\IEEEauthorrefmark{1}\IEEEauthorrefmark{2}\orcidlink{0000-0001-9483-7562}}
    \IEEEauthorblockA{%
        \IEEEauthorrefmark{1}Masaryk University, Faculty of Informatics, Brno, Czech Republic}
    \IEEEauthorblockA{%
        \IEEEauthorrefmark{2}University of Bergen, Department of Informatics, Bergen, Norway}
}

\maketitle

\begin{abstract}
Getting acquainted with a large codebase can be a daunting task for software developers, both new and seasoned.
The description of a codebase and its development should be the purpose of its documentation.
However, software documentation, if it exists at all, is usually textual and accompanied only by simple static diagrams.
It is also time-consuming to maintain manually.
Even an API reference, which can be generated automatically from the codebase itself, has many drawbacks.
It is limited to what it can extract from the codebase, is cumbersome to navigate, and fails to capture the interwoven nature of code.
We explore an alternative approach centered around a node-link diagram representing the structure of a codebase.
The diagram is interactive and filterable, providing details on demand.
It is designed for automation, \rev{relying on static analysis of the codebase, and thus produces results quickly and offers} a viable alternative to missing or outdated documentation.
To evaluate this approach, we implemented a prototype named Helveg that is able to analyze and visualize C\# code.
Testing with five professional programmers provided feedback on the approach's benefits and challenges, which we discuss in detail.

\end{abstract}



\section{Introduction}


Nearly every piece of software eventually reaches a point when it should be documented.
This is because technical documentation is the most sought-after learning resource~\cite{so_survey}.
Well-written documentation is the hallmark of good open-source projects and can be critical for their adoption~\cite{good_docs}.
Unfortunately, real-world documentation is often incomplete or confusing~\cite{oss_survey} due to time constraints.

Nowadays, software documentation is typically written in a descriptive textual form and is hosted on the web to be easily accessible.
The text is frequently enhanced with code snippets and occasionally with simple static charts and diagrams.
Developers often use automatic generators to extract special comments and metadata directly from the source code, reducing the hassle of writing documentation.
Such generators typically output a page or an entry for every significant code element (e.g., a type or a function).
The documentation generated this way is often collectively described as an \textit{API reference}.
The automatic creation of an API reference saves precious development time and makes it easier to keep documentation up-to-date with the code.
However, an API reference has many drawbacks.
It is limited to code metadata and the contents of \textit{documentation comments}.
These comments are written manually by the developers and, thus, are often omitted.
The navigation of an API reference usually relies on hyperlinks, full-text search, and a table of contents mimicking the module hierarchy.
Thus, this limited navigability makes it cumbersome to look for specific arrangements of code elements like types bound together through composition.
Therefore, authors often use multiple writing methods in practice, combining their advantages.
They write manually about key concepts spanning multiple code elements and delegate the details of each code element to the automatic API reference generators.

While documentation can benefit greatly from using visualization~\cite{whiteboard, rost_survey}, its use is currently limited.
It is usually confined to diagrams that describe the project's architecture, explain its type hierarchies, or compare the project to its competitors.
While these visualizations are undoubtedly helpful, they are seldom interactive, are often created manually, and usually act only as a support for their surrounding text.
Therefore, they make writing documentation even more demanding because authors need knowledge of a graphic editor or a tool like GraphViz~\cite{graphviz}.
Even diagrams, such as those generated by Doxygen~\cite{doxygen}, are only static images and thus do not utilize the interactive potential of digital platforms.

To see the challenges with software documentation, consider a scenario where a new developer is assigned to a project. 
Navigating the existing codebase can be challenging without well-maintained documentation or the help of a senior colleague---both of which are frequently unavailable.
Often, the newcomers are left only with an API reference that offers a detailed view of each code piece but not a high-level overview of the entire project.
As the newcomers are unfamiliar with the codebase, they cannot effectively utilize full-text search.
Consequently, they are left to browse the entire documentation, hoping to stumble upon the information they need.
This process is inefficient and often frustrating.
Alternatively, they can generate a static diagram showing the whole project.
However, these diagrams typically result in unreadable hairballs or, conversely, omit critical details.

In this paper, we explore an alternative to a classical API reference.
One where the documentation can still be generated automatically but is navigated through an interactive node-link diagram.
This diagram enables a high-level, exploratory analysis of a codebase without trudging through its implementation details.
However, it is also flexible enough to provide these details on demand through filtering and interaction.

To test this concept, we built a prototype tool called Helveg.
Helveg can analyze a codebase written in the C\# language and automatically output a web application with an interactive diagram representing the structure of the codebase.

\pagebreak
In summary, the main contributions of this paper are:

\begin{itemize}
    \item An alternative concept to API references for exploring codebases. The approach is based on an interactive diagram providing a high-level overview as well as details on demand.
    \item An open-source prototype implementation of the concept and a data miner for C\# codebases. Their source code is available at \url{https://gitlab.com/helveg/helveg}.
    \item An evaluation of the effectiveness of the presented concept with professional C\# developers.
\end{itemize}

\vspace{0.1em}


\section{Background and Related Work}


Documentation is connected to the topic of program comprehension, which many software visualization (SV) methods aim to improve.
These methods often focus on visualizing the structure of a program's source code, the program's behavior at runtime, or its evolution.
Since this paper proposes a documentation approach that concerns itself only with the static structure of a codebase, we further delve into techniques that do the same.
We also compare many existing open-source and commercial programs that enhance software documentation or visualize code structure.
Regarding SV as a whole, we refer to literature reviews by Chotisarn et al.~\cite{slr} and Bedu et al.~\cite{slr_2}.

\subsection{Visualization of Code Structure}


A simple way of visualizing a large amount of source code is to scale it down.
This is how, in the code-map metaphor~\cite{code_map_review}, code lines become colored, elongated rectangles.
It can be used to visualize length, age, programming language, and other code file metrics.
SeeSoft by Eick et al.~\cite{seesoft} is a well-known example of this method.
This technique continues to be researched in recent years, as proven by CodePanorama by Etter and Mehta~\cite{codepanorama}, and it is even included in the popular Visual Studio Code editor as its Minimap feature~\cite{vscode_minimap}.

A wholly different class of code structure SV methods is based on real-world metaphors.
One such metaphor is the software city, in which codebases transform into 3D cities, where buildings represent units of code, such as classes, and their dimensions and colors correspond to code metrics.
The idea has given rise to diverse implementations such as Software World~\cite{software_world}, CodeCity~\cite{code_city}, CodeMetropolis~\cite{codemetropolis}, and BabiaXR~\cite{babiaxr}.
The city metaphor is further extended to islands populated by cities in IslandViz by Schreiber and Misiak~\cite{islandviz}.
While the metaphors are intuitive, they frequently omit relationships between code elements~\cite{city_metaphor_overview} and suffer from occlusion issues common to all 3D visualizations~\cite{munzner2014visualization}.

Apart from code maps and software cities, there is a plethora of 2D charts and diagrams.
These are designed for diverse purposes, spanning from visualization of software architecture and dependencies between modules to reverse engineering and optimization.
To name a few, Lanza and Ducasse~\cite{polymetric_views} presented polymetric views, a system with multiple views capable of displaying diverse metrics.
Gutwenger et al.~\cite{uml_approach} displayed UML class diagrams in a new way involving the minimization of crossings, orthogonal edges, uniform edge direction within type hierarchies, and other desirable criteria.
In E-Quality by Erdemir et al.~\cite{equality}, the codebase is displayed as a node-link diagram, where nodes are glyphs representing quality metrics.
Müller and Zeckzer~\cite{recursive_disk} also take a glyph-based approach, in which the structure and relations in a codebase become a circular treemap.
There are also methods utilizing rectangular and circular treemaps, like Git-Truck~\cite{git-truck} by Højelse et al.

A common problem all of these approaches must tackle is scalability~\cite{slr_2}.
Even relatively small software projects involve a large amount of data.
Thus, visualizations must be equipped with some kind of filtering or aggregating mechanism to be readable.
The problem with scalability also applies to performance, as SV tool makers have to consider the complexity of their analysis and rendering algorithms.

To summarize, SV research in the area of code structure and software architecture offers a diverse group of techniques.
Our attempt shares some of the challenges with these techniques.
\rev{The scalability issue, in particular, is one we encounter as well.}
However, we also face additional challenges stemming from our goal to offer a better alternative to API references.

\subsection{Existing Tools}
\label{sec:existing-tools}



Many existing tools can be used to visualize the structure of source code or software architecture.
In this section, we mention several examples and group them based on features relevant for use in software documentation.
For a more extensive comparison, see Supplementary Material~\cite{sm}.

\textbf{Conformance to standards.}
Architecture-visualizing tools typically produce diagrams that adhere to an existing standard, most commonly to UML class diagrams~\cite{uml}.
The advantage of this approach is that users are often already familiar with how the data is visualized.
Many tools can generate UML class diagrams from source code, such as the UML Diagrams plugin~\cite{jetbrains_uml} for JetBrains IDEs and Visual Paradigm~\cite{visual_paradigm}.
Apart from UML, there is also the less strict C4~model~\cite{c4} along with the Structurizr tool~\cite{structurizr}.
However, both UML and C4 were designed with static images in mind that can be, for instance, drawn on a whiteboard or printed.
These tools usually do not take advantage of the potential for interactivity present in digital media, like the ability to expand and collapse nodes or provide detailed metadata on demand.
Therefore, numerous architectural visualizers do not follow any standard, allowing them to include more language-specific details.
These include tools like the Visual Studio Code Map~\cite{code_map} (unrelated to the code-map metaphor) and all tools employing the city metaphor or a similar 3D approach.

\textbf{Code analysis / Automation.}
A significant distinguishing factor between the tools is whether they perform any \textit{code analysis}.
In the context of this paper, we define \textit{code analysis} as any data mining or processing step involving language parsing and semantic analysis.
For example, tools like Structurizr~\cite{structurizr} and D2~\cite{d2} are text-to-diagram utilities, turning a manually written description into a corresponding diagram.
On the other hand, NDepend~\cite{ndepend} and JArchitect~\cite{jarchitect} measure code quality and thus, apart from visualizing a codebase, also heavily analyze it.
Similarly, API reference generators, such as Doxygen~\cite{doxygen} and Docfx~\cite{docfx}, can \rev{analyze a software project} and generate static diagrams depicting inheritance hierarchies and attach them to relevant pages.
From the user's perspective, code analysis is the difference between describing a codebase manually and simply passing a path to a tool as an argument.
Tools performing code analysis can run automatically, and automation is a key requirement for providing an alternative to API references.

\textbf{Interactivity.}
While a static diagram of a small codebase can be readable, it often becomes cluttered as the codebase grows.
This issue can be largely avoided by providing details-on-demand features---letting the user interact with the visualization and choose what they want to focus on~\cite{shneiderman}.
For instance, the Emerge tool~\cite{emerge} features context menus, filtering, highlighting, and a force-directed graph layout algorithm.
However, Emerge focuses more on code quality than documentation and, for example, disregards documentation comments.

\textbf{Web support.}
Any alternative to an API reference must be ready for the web environment.
This is because software documentation tends to be hosted online, where anyone can access it.
Although many tools are interactive and can generate diagrams \rev{from data obtained by code analysis}, they are distributed as binary programs with a graphical interface.
Therefore, they lose their interactivity when their diagrams are exported as static images.
These tools also often cannot run during automated builds, making them difficult to use as alternatives to an API reference.
Commercial web platforms, such as CodeSee~\cite{codesee} and Swimm~\cite{swimm}, offer a web experience, but they focus on developer collaboration and making documentation maintainable rather than on code analysis.
There are also open-source tools like Emerge~\cite{emerge} and DependenSee~\cite{dependensee} that output their diagrams as self-contained web applications, which can be placed next to the documentation as is.
However, while they demonstrate their strength in embeddability, the scope of their analysis is limited to module dependencies.
The final mention goes to the Obsidian note-taking tool---a program for managing Markdown notes connected by hyperlinks~\cite{obsidian}.
It also features a node-link diagram displaying the connections between files.
Despite not performing code analysis, the Obsidian development team utilized it for their API reference.

\thematicbreak

In summary, none of the mentioned tools provides a combination of features that would make it a suitable replacement for an API reference.
They either cannot be automated, are not interactive, or cannot be embedded into existing documentation hosted on the web.


\section{Requirement Analysis}

This paper explores the assumption that interactive visualization of a codebase's structure can improve documentation and facilitate exploratory analysis.
Our solution was designed around this assumption while considering several requirements drawn from two sources.
The first source was a survey among professional programmers, programming tutors, and computer science students conducted within the scope of a master thesis~\cite{mgr}.
The other source was our comparison of existing tools and their strengths and weaknesses, which we summarized in the previous section.
Based on the collected information, we settled on the following goals.

\begin{enumerate}[label={\textbf{R\arabic*:}}, ref=R\arabic*, leftmargin=*]
    \item\label{req:familiar}\label{req:first} Output a visualization that feels familiar.
    \item\label{req:analyze} Generate the visualization automatically \rev{using data from} source code \rev{analysis}.
    \item\label{req:interact} Allow user interactions to enable exploration and filtering of the codebase on different levels of granularity.
    \item\label{req:embed}\label{req:last} Output visualization in a format that can be embedded into a website.
\end{enumerate}

Apart from the high-level requirements listed above, we also had to decide on the technical aspects of the implementation, as we needed to test the concept in practice.
In particular, we needed to consider what programming languages to support, as this decision is important for data mining algorithms.
In this regard, we identified C\# as an ideal target language for our prototype implementation.
As our main goal was to test the visual representations, we decided to support only one programming language.
The advantages of C\# are its large user base~\cite{so_survey}, numerous projects to visualize, and an API for directly inspecting the compiler's semantic model.
However, we are convinced that the concept presented in this paper also applies to other object-oriented programming languages.
The following section summarizes the specifics of the C\# programming language, whose knowledge is crucial for understanding our solution and its components.


\section{C\# Language and its Specifics}
\label{sec:csharp}


Our visualization design and its prototype implementation assume familiarity with object-oriented programming (OOP) and particularly the C\# programming language.
This section explains several C\# terms, constraints, and guarantees relevant to this paper.

\textbf{The .NET platform.}
The C\# language is a part of the .NET platform, which envelops several other languages, runtimes, and related tooling.
There are three relevant pieces of tooling: Roslyn, NuGet, and MSBuild.
The .NET Compiler Platform (``Roslyn'')~\cite{roslyn} is the C\# compiler, providing many \textit{code analysis} APIs, which allow others to build tools capable of understanding C\#'s syntax and semantics.
NuGet~\cite{nuget} is .NET's package manager facilitating the resolution and acquisition of external dependencies.
Finally, Microsoft Build Engine (MSBuild)~\cite{msbuild} is the build system tying Roslyn and NuGet into a pipeline that transforms a description of a C\# project, its source code, and dependencies into a binary artifact called \textit{an assembly}.

\textbf{The project system.}
MSBuild works with \textit{projects} and \textit{solutions}.
C\# projects define what NuGet packages are downloaded, what C\# files are fed to Roslyn, and what other steps are performed as part of the build process.
Visual Studio Solutions are then simply sets of projects.
MSBuild and NuGet prevent projects and packages from having circular dependencies.

\textbf{The type system.}
Object classes are the core concept of OOP.
In C\#, classes are just one kind of \textit{types} the language has to offer.
There are five \textit{type kinds} in total: classes, structures, enumerations, interfaces, and delegates.
All types must be declared inside a hierarchical \textit{namespace}.
The top-level namespace is the \textit{global namespace}.
Types may contain \textit{type members}, which fall into one of four categories: fields, methods, properties, and events.
Both types and their members have an \textit{accessibility}---a degree to which other pieces of code can interact with them.
For example, \textit{public} methods can be called from anywhere, while \textit{private} ones are accessible only within their declaring type.
For a more thorough description of C\# types, see .NET's Common Type System~\cite{dotnet_cts}.

\textbf{The static modifier.}
Class types and type members can have the \texttt{static} modifier.
This modifier has special importance since \texttt{static} members are accessible without an instance of their declaring type.
Static classes are then types that may contain \textit{only} static members.
For a C\# programmer, a static class is essentially a container for global variables and functions.
This is important since, in .NET, variables and functions cannot be declared outside of types.

\textbf{The entity abstraction.}
Roslyn, NuGet, and MSBuild each have a different abstraction of the codebase they act upon.
For our purposes, we unify them into our \textit{entity abstraction}.
Each \textit{entity} has a name and an \textit{entity kind}, which can be, for example, \textit{solution}, \textit{project}, \textit{package}, \textit{type}, or \textit{method}.
Entity kinds form a hierarchy since it can be said that a solution contains a project, which contains a type, which declares a method, and so on.
Some entity kinds are further specialized.
For example, types have a \textit{type kind} (e.g., class, struct, enum, etc.), and methods have a \textit{method kind} (e.g., constructor, getter, operator, etc.).


\section{Visualization Design}
\label{sec:design}

Following Munzner's data visualization framework~\cite{munzner2014visualization}, we iterated and enhanced our visual alternative to an API reference.
The main influences on the final visualization design and interactions were the \ref{req:first}--\ref{req:last} requirements, team discussions, and hands-on experience with our prototypes.

We considered several visual metaphors, including node-link diagrams, UML diagrams, and 3D cities.
Our goal was to create a visual representation that could provide both an overview of the entire project and details of individual code elements (see \ref{req:interact}).
We disregarded standardized UML diagrams because they would require different visual representations for the overview and the details, thereby increasing complexity and the learning curve, which was unacceptable given our aim to simplify documentation navigation.
Similarly, we dismissed the 3D city metaphor in favor of simplicity and effectiveness, as 3D visual channels are less effective compared to 2D channels~\cite{munzner2014visualization}.

\begin{figure}[tb]
  \centering
  \includegraphics[width=1.0\linewidth]{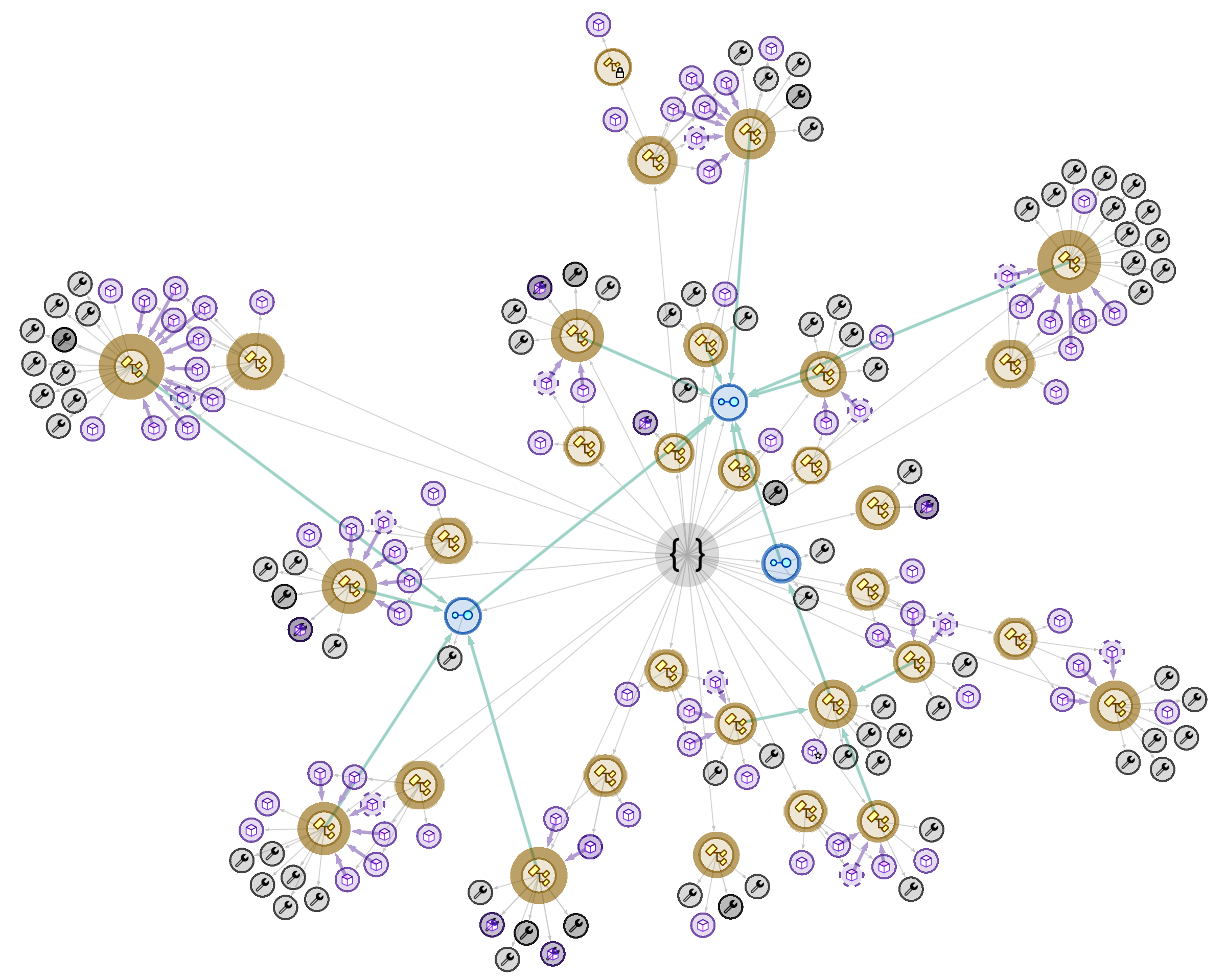}
  \caption{
    A sample diagram of a namespace with several types (yellow and blue nodes).
    It contains several relations: \texttt{declares} (gray edges), type inheritance (teal edges), and method return types (purple edges). Labels were omitted.
    }
  \label{fig:example} 
\end{figure}

Therefore, we opted for a node-link diagram, representing the codebase as a graph, i.e.,~a set of nodes and edges.
Nodes correspond to \textit{C\# entities}, and edges represent relationships between them. Figure~\ref{fig:example} shows an example of this diagram.
While node-link diagrams were already used for visualization of codebases before~\cite{code_map, obsidian}, to the best of our knowledge, they were never explored as primary means of navigation within an API reference.
We can also benefit from the fact that node-link diagrams are a well-known and generally understood concept, which enhances the adoption of the solution.

\subsection{Glyphs}
\label{sec:glyphs}

\begin{figure}[tb]
  \centering
  \includegraphics[width=0.45\linewidth]{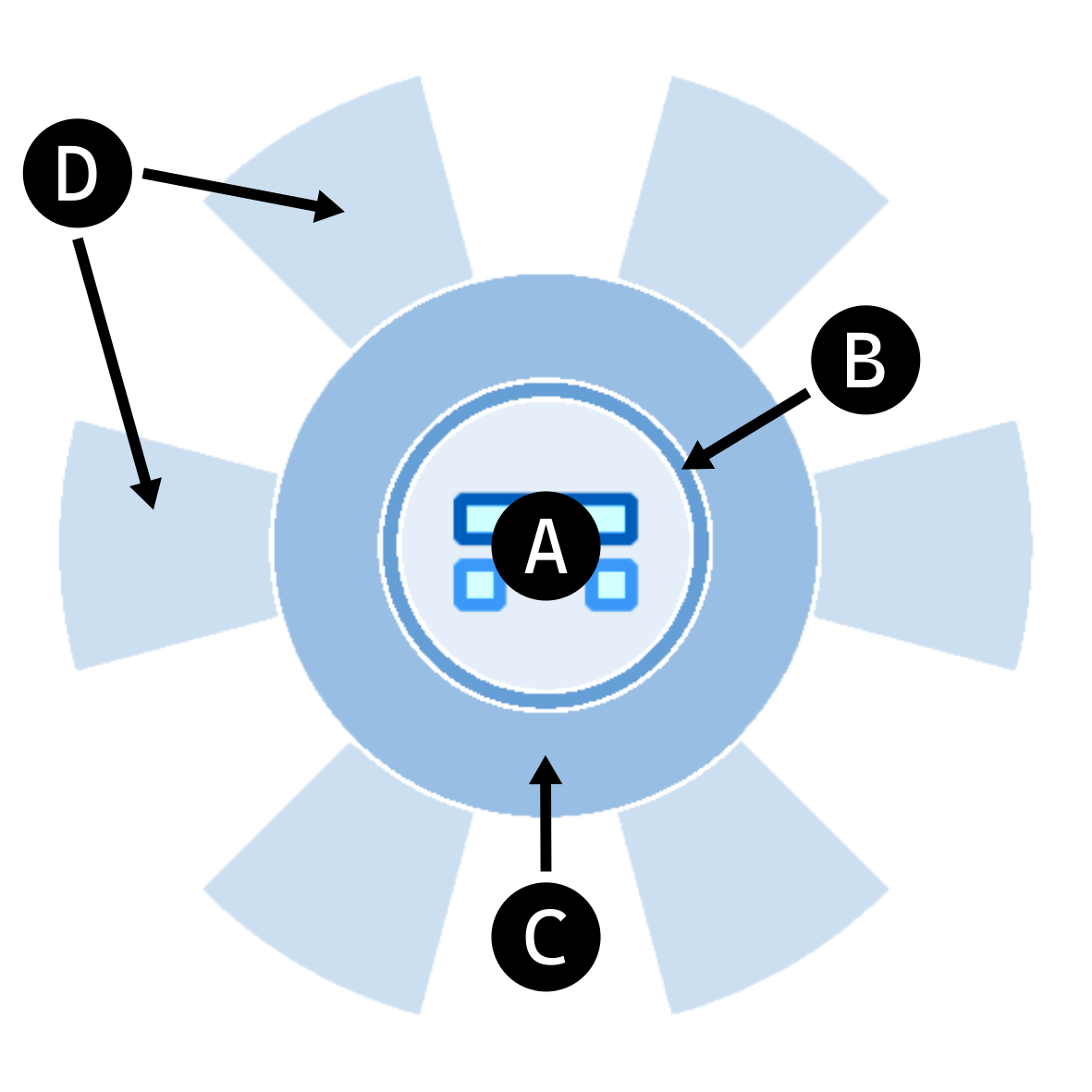}
  \caption{\label{fig:struct-glyph} A glyph-node representing a C\# type: (A)~The icon of a C\# struct. (B)~The innermost outline indicates that this type is non-static. (C)~The middle, solid outline represents the amount of non-static members. (D)~The outer, dashed outline stands for the amount of the type's static members.}
\end{figure}

The nodes of the node-link diagram are glyphs---compound visual elements symbolizing several data attributes at once (see Figure~\ref{fig:struct-glyph}).
We have decided to use glyphs as they provide an ideal compromise between space-efficiency and understandability~\cite{glyphs}.
All nodes are circular, as is typical for node-link diagrams.
Circles can also be arranged in a way that minimizes overlap and maximizes the use of space in the diagram, which is particularly useful for complex networks.
The node radius is calculated based on several factors.
First, the represented entity's kind determines the fixed starting radius.
For example, the starting radius of a project node is larger than that of a type node.
For type entities, the starting radius is then increased by their member count.
The radius is then scaled, employing either a linear, logarithmic, or square-root scaling method.
Users have the flexibility to configure the scaling method, which can be helpful when the diagram contains excessively large nodes.

\begin{figure}[t]
  \centering
  \includegraphics[width=\linewidth]{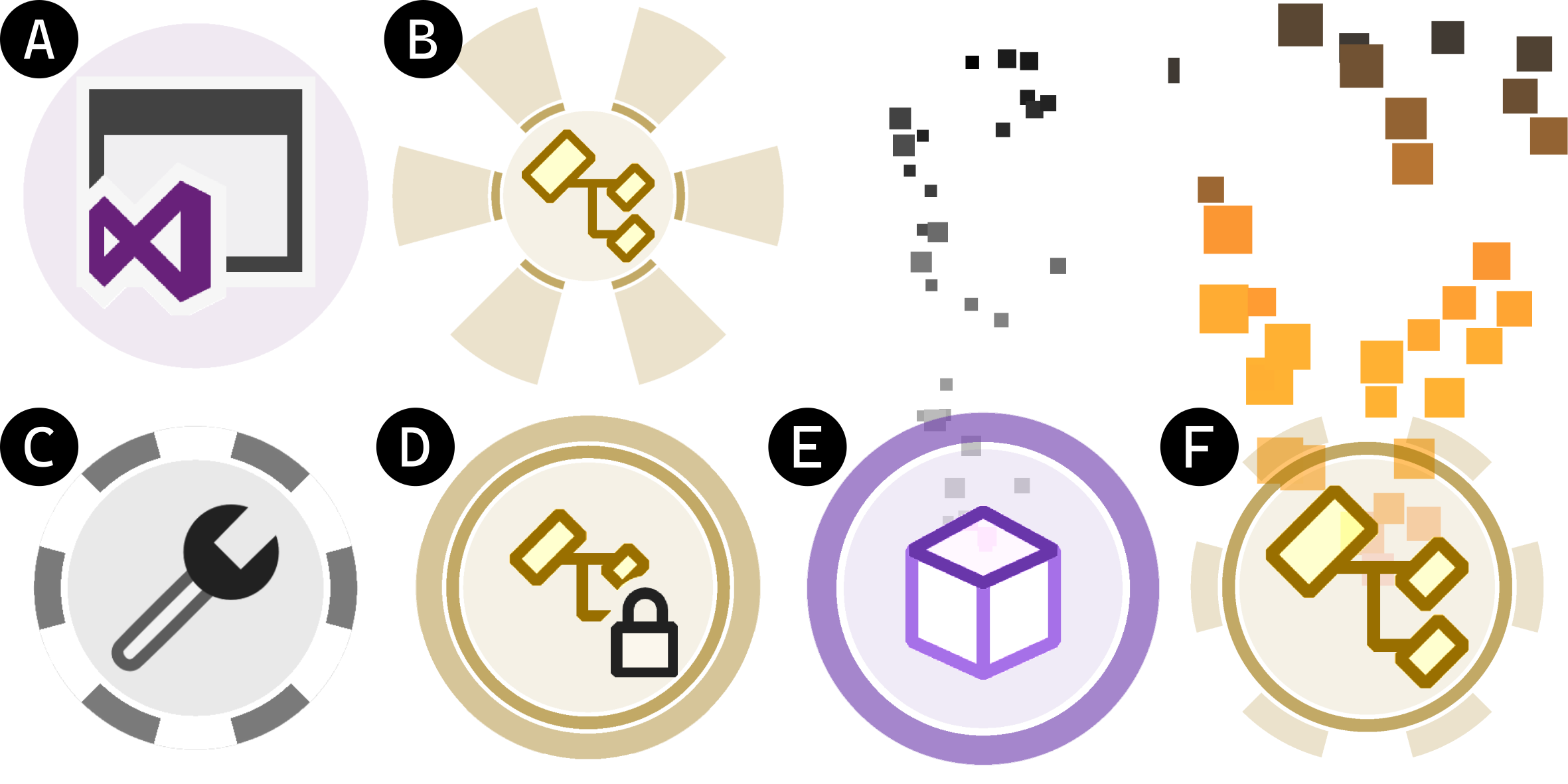}
  \caption{\label{fig:glyphs} Examples of glyph-nodes: (A)~A~solution. (B)~A~static with only static members. (C)~A~static property. (D)~A~non-static private class with a small amount of non-static members. (E)~A~method with a compiler warning. (F)~A~class with a compiler error.}
\end{figure}

Each glyph has an icon in the middle representing its \textit{entity kind} (Figure~\ref{fig:struct-glyph}-A).
In the case of types, the icon represents their \textit{type kind} as it is more specific (see Section~\ref{sec:csharp}).
Figure~\ref{fig:struct-glyph}, for instance, represents a \texttt{struct} type, whereas Figure~\ref{fig:glyphs}-B is a \texttt{class}.
These icons are taken from the Visual Studio (VS) image library~\cite{vs_icons}, courtesy of Microsoft, which allows their use in third-party tools.
They were chosen intentionally since VS is the most common IDE among C\# programmers~\cite{csharp_survey}, which aligns with~\ref{req:familiar}.
The icon also sets the node's tint to strengthen its kind's distinctiveness.

Another highly relevant information is \textit{accessibility} of types and their members. 
Its value can range from \texttt{public} to \texttt{private} with many nuanced shades in between~\cite{ecma-335}.
If the represented entity is anything other than \texttt{public}, the glyph contains a smaller icon in the lower right corner of the kind icon.
For example, Figure~\ref{fig:glyphs}-B is a \texttt{public} type since it has no icon in its corner.
However, Figure~\ref{fig:glyphs}-D is \texttt{private} because its corner icon depicts a lock.

Type glyphs also have up to three \textit{outlines}, as visualized in Figure~\ref{fig:struct-glyph}.
The innermost outline (Figure~\ref{fig:struct-glyph}-B) represents the \texttt{static} modifier.
If the outline is dashed, the entity is \texttt{static}; a solid outline corresponds to a non-\texttt{static} entity.
The two outer outlines are only present on type nodes and symbolize the number of members the type declares.
The middle outline (Figure~\ref{fig:struct-glyph}-C) is always solid, and its width represents the number of non-\texttt{static} members.
Similarly, the outermost outline (Figure~\ref{fig:struct-glyph}-D) is always dashed, and its width symbolizes the number of the type's \texttt{static} members.
The order of the outlines is always the same.

The visual element of node outlines has been chosen deliberately, as we believe it adequately represents the information and its relevance to the type.
To explain, the presence of the \texttt{static} modifier is a \rev{detail} that is the closest to the type and, thus, deserves to be depicted as the innermost outline.
Similarly, non-static members are more closely tied to the type than static members and, therefore, are closer to the node's center.
We further emphasize this closeness by decreasing the saturation of each outline from the node's center outward.

A subset of these rules also applies to type members---nodes with the \textit{property}, \textit{field}, \textit{event}, or \textit{method} entity kind.
On these nodes, the accessibility icons and innermost outline representing the \texttt{static} modifier are also displayed.
However, their size is constant and smaller than that of any type node.

Take, for example, the nodes from Figure~\ref{fig:glyphs}:
\begin{enumerate}[label={\helcircled{\Alph*}}, leftmargin=*]
    \item is the solution node and the root of the codebase, which is immediately clear from its icon from VS.
    \item represents a \texttt{static} class with only \texttt{static} members.
        This is made clear by the dashed innermost outline, the presence of a dashed outer outline, and the complete absence of a solid outline.
    \item is a static property, given by the property icon and the dashed outline.
    \item is a non-\texttt{static} class with only a handful of instance members. It is also \texttt{private} since it has the lock icon.
\end{enumerate}

The glyph-nodes may also exhibit one of two animated effects.
These effects symbolize Roslyn diagnostics---compiler warnings and errors\rev{---which point out problematic pieces of code. They allow a team of developers to review issues within the codebase alongside the documentation.}
Specifically, if a node has an error diagnostic attached to it, it appears to be on fire (Figure~\mbox{\ref{fig:glyphs}-F}).
In the case of compiler warnings, the node emits a mere column of smoke instead (Figure~\ref{fig:glyphs}-E).
These effects were chosen as fire and smoke are typically treated with caution, similar to compiler diagnostics.
We also chose to animate them, as that makes them more noticeable, which is fitting due to the diagnostics' importance.

\subsection{Entity Relationships}

Edges of the node-link diagram represent relationships between C\# entities.
Each edge is defined by its source and destination node as well as a \textit{relation}---a set of edges---it is part of.
There are multiple relations between nodes since we have decided to visualize more than just ``dependencies'' among the entities to represent the codebase accurately.

The \texttt{declares} relation mirrors the \textit{entity kind hierarchy} (see Section~\ref{sec:csharp}).
As such, it most clearly represents the codebase structure.
For example, in this relation, a \textit{solution} has outgoing edges into \textit{projects}, as solutions declare projects.
Similarly, a \textit{namespace} is linked to \textit{types} it contains, which are connected to \textit{methods} declared by them, and so on.
However, the \texttt{declares} relation alone offers no advantage over a table of contents commonly found in API references.
The real benefit of the node-link diagram lies in its ability to represent multiple relations at once.
Other relations symbolize entity associations such as type inheritance (\texttt{inheritsFrom}), types of fields/properties/parameters (\texttt{typeOf}), and project dependencies (\texttt{dependsOn}).
For an exhaustive list of the visualized relations, see Supplementary Material~\cite{sm}.

To distinguish between relations when depicted simultaneously, each has a different color and line weight applied to their edges.
Each color is customizable, and each relation can be disabled if it is undesirable.
This way, the amount of clutter caused by an excessive amount of edges and their crossings can be minimized.
Following this mindset, we display only the \texttt{declares} edges between the nodes by default. 

\subsection{Scalability}
\label{sec:scalability}

A typical problem that node-link diagrams face, especially when used to explore large datasets, is visual clutter and overlapping, which can obscure important relationships and make the diagram difficult to interpret.
A common solution found in literature~\cite{schulz2011treevis} is to aggregate or filter the unimportant nodes. 
In this paper, we have primarily focused on filtering (see Section~\ref{sec:filtering}) as it was the most requested feature in the initial user survey~\cite{mgr}.
Therefore, we wanted to explore to what extent filtering would suffice as the primary means of improving the diagram's readability.

We utilize aggregation to some extent by allowing users to expand and collapse nodes at will (see Section~\ref{sec:interactivity-layout}).
However, one could argue that this is a form of localized filtering based on code hierarchy, as we do not visually communicate the cumulative properties of the aggregated nodes.
This was a deliberate design choice, as we wanted to avoid bias in the evaluation of the filtering features.

\subsection{Filtering and Highlighting}
\label{sec:filtering}

As explained above, we strive to provide the most flexible search and highlight feature we can.
Therefore, we have decided on three search modes: \textit{full-text} search through the entity names, \textit{regex} doing the same but using a regular expression, and the \textit{JavaScript (JS)} mode allowing the user to write their own search logic with a custom JS filter.
In any mode, the user has a choice to either \textit{highlight} the matching nodes, graying out the rest, or \textit{isolate} the matches and remove the remainder from the current state of the diagram.

Among the search modes, \textit{JS} is the most flexible one.
It enables the user to filter nodes based on any piece of entity metadata and not just their name.
For instance, one might want to know which nodes contain compiler errors, possess a number of members exceeding a certain threshold, or contain documentation comments with a certain keyword.
All of these queries can be made using the JS search mode.

\subsection{Interactivity and Layout}
\label{sec:interactivity-layout}

Apart from filtering, our approach relies on interactivity to be readable even as the diagram grows (see \ref{req:interact}).
By default, the diagram shows only the solution and project nodes, with the rest of the codebase hidden.
The user can use filtering or manual interaction to \textit{expand} nodes and show their children, thus shifting focus to a specific area of the codebase.
Likewise, nodes can be \textit{collapsed} to unclutter the diagram and gain a more high-level view.
Parts of the diagram can also be \textit{removed} from the current view.
These removed parts are then excluded from filtering and further interaction until the diagram is refreshed.
Finally, users can select any node they are interested in and inspect the represented C\# entity in a side panel (see Figure~\ref{fig:ui}-D).

The layout of the node-link diagram needs to respond to these interactions and remain readable.
However, users can still reposition nodes manually if they are dissatisfied with the automatic layout.


\section{Prototype Implementation}

\begin{figure*}[tb]
  \centering
  \includegraphics[width=1.0\linewidth]{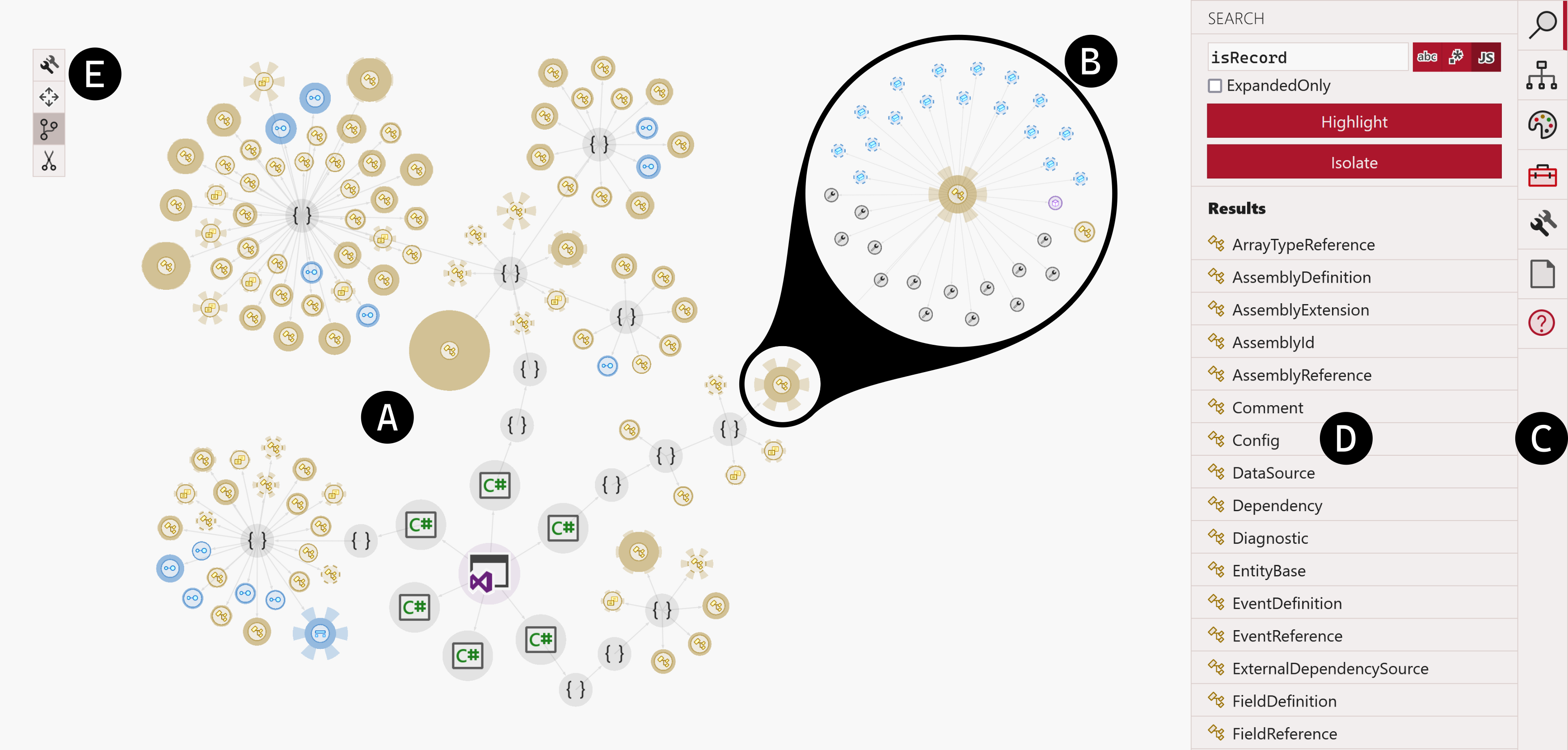}
  \caption{Helveg's user interface: (A)~The diagram component. (B)~A type node before (smaller circle) and after expansion (larger circle) using the Toggle tool. (C)~Dock allowing to switch between panels. (D)~A panel containing search controls. (E)~Toolbox with tools to directly modify the diagram.}
  \label{fig:ui} 
\end{figure*}

To test our alternative API reference, we built a prototype named Helveg.
Helveg consists of three main components: a data miner, the interactive diagram itself, and a JavaScript Object Notation (JSON) schema used to communicate between the first two components.
The C\# codebase data miner is implemented in C\# as a console application so that it can be readily run in a server environment (see \ref{req:analyze}).
The diagram is then implemented in TypeScript (TS) and wrapped in a web-based user interface (UI) (see Figure~\ref{fig:ui}).

The console application implements a pipeline that (1) analyzes a C\# codebase, (2) serializes its abstraction into JSON, and (3) outputs a web application containing the interactive diagram.
The web application can optionally be assembled into a single HTML file.
This form of deployment has the advantage of not requiring a server to run on the user's machine~\cite{cors-not-http}.

The rest of this section describes the most important aspects of Helveg's implementation.
For additional details, see Supplementary Material~\cite{sm}.

\subsection{Data Mining}
\label{sec:data-mining}

As mentioned, the console application analyzes the visualized C\# codebase---performs data mining (see \ref{req:analyze}).
To be more specific, the data miner implements our unified C\# entity abstraction (see Section~\ref{sec:csharp}).
Therefore, it integrates with the APIs of Roslyn, MSBuild, and NuGet.
None of the official APIs were designed with visualization in mind.
Thus, the data miner must perform substantial additional processing and work around the limitations of the APIs.

One of the main issues we encountered was the unique identification of C\# entities.
This is required so that the entities can be serialized to JSON and passed to the diagram.
This issue was surprisingly difficult to solve due to C\#'s \textit{method overloading} and \textit{generics}, which prevented us from using fully-qualified names of the entities since they may contain circular references.
Our implementation avoids these circular references by relabeling the entities with a unique, hierarchical, numeric token.
These tokens are essentially number sequences, encoding the path from a \texttt{declares} root to the identified entity.
The numeric tokens also allow the data miner to analyze the codebase in multiple threads and make the diagram's layout more consistent across multiple tool runs.

Each entity also has \textit{diagnostics}, \textit{comments}, and metadata appropriate for its entity kind.
Diagnostics represent compiler warnings, errors, and hints.
This allows Helveg to visually represent codebases that are not compilable and highlight the entities responsible for the errors.
Similarly, the tool extracts documentation comments, ensuring that any documentation within the codebase is available in the final web application.

\subsection{The Diagram}

Helveg's visualization, the node-link diagram described in Section~\ref{sec:design}, is implemented in TypeScript and transpiled into JavaScript, which can run in any modern web browser (see \ref{req:embed}).
It uses the Graphology~\cite{graphology} library, which provides the necessary graph theory abstractions, and the Sigma.js~\cite{sigma} library, which renders the node-link diagram itself.
While these two libraries form the basis of the functionality, Helveg modifies Sigma's layout behavior and adds its own WebGL shaders that render the glyph-nodes.

Helveg relies on two graph layout algorithms---TidyTree by Reingold and Tilford~\cite{tidytree}, and ForceAtlas2 by Jacomy et al.~\cite{forceatlas2}.
TidyTree positions the nodes in a circular dendrogram.
It executes almost instantaneously, but it requires the underlying relation to be a tree and thus is used only on \texttt{declares}.
ForceAtlas2 is an iterative, force-directed graph layout algorithm, which \rev{runs} after TidyTree has been applied.
Compared to TidyTree, it \rev{is not limited to trees} and is able to respond to user interaction, but its convergence may \rev{be slow}, depending on the number of nodes.
Helveg relies on open-source implementations of these algorithms: TidyTree from the D3 library~\cite{d3} and ForceAtlas2 from Graphology~\cite{graphology}.

The diagram is encapsulated by a UI implemented in Svelte~\cite{svelte}.
It consists of several distinct parts besides the diagram itself.
The \textit{Dock} (Figure~\ref{fig:ui}-C) is a vertical bar housing several tabs.
It switches between panels, which occupy the same space (Figure~\ref{fig:ui}-D) on the screen.
Panels \rev{may contain the properties and documentation comments of the currently selected node, search results,} or configuration options.

By default, the user selects nodes by clicking on them with the left mouse button.
However, this behavior can be changed by selecting another tool in the \textit{Toolbox} (Figure~\ref{fig:ui}-E).
These tools correspond to the interactions described in Section~\ref{sec:interactivity-layout}.

The design of the UI draws inspiration from commercial software that the users are likely familiar with already.
We believe exploiting this familiarity~(\ref{req:familiar}) makes the UI more intuitive and easy to use.
The Toolbox is inspired by toolbars on the left side of the screen, which are present in many graphic editors.
Similarly, the Dock and its Panels are heavily inspired by those in Visual Studio Code.

Most of the diagram's functionality can be customized to suit the user's needs.
All of the configuration options are contained in the panels.
For example, in the \textit{layout} panel, the user can use the provided checkboxes to select the entity kinds and relationships they are interested in.
Corresponding nodes and edges will be laid out once the \textit{Refresh} button is clicked.
The same panel houses controls of the TidyTree and ForceAtlas2 layout algorithms.
All of the panels are described and pictured in Supplementary Material~\cite{sm}.


\section{Helveg User Study}

The functionality of the Helveg prototype was evaluated through an exploratory user study.
The study consisted of semi-formal testing sessions with five software developers (evaluators \textbf{E1}--\textbf{E5}), whom we recruited by word of mouth.
They all have at least five years of professional C\# experience at a software company and regularly encounter software documentation.
The testing sessions were about an hour long, during which the participants gave their feedback verbally and via a questionnaire.
A summary of each interview and the testing samples are attached to Supplementary Material~\cite{sm}.

We wanted to direct the participants' attention to all of Helveg's features, so we provided them with a sample diagram of a project and several tasks.
The project was KAFE~\cite{kafe}, an open-source web archive of multimedia implemented in C\#.
None of the participants had encountered KAFE before their interview.
On the other hand, the interviewer was one of KAFE's developers and thus could easily assess the accuracy of the testers' observations.
The participants were also deliberately not introduced to Helveg to simulate the behavior of users who had only just discovered our visualization.
They were, however, \rev{told what Helveg's purpose is and} given a \rev{hint} if they asked for it or got stuck.
The tasks given to them were the following questions paired with assignments:

\begin{enumerate}[label={\textbf{T\arabic*:}}, ref=\textbf{T\arabic*}, leftmargin=*]
    \item\label{task:projs} What are the most important projects of the KAFE solution?
        \textit{Identify what the solution's projects are and what their purpose is.}
    \item\label{task:deps} What external dependencies do the projects have?
        \textit{Determine on which packages the projects depend.}
    \item\label{task:smells} Are there any issues with the codebase?
        \textit{Identify any errors, code smells, or other issues with the codebase.}
    \item\label{task:diff} What are the major changes between the compared versions of KAFE?
        \textit{Try to deduce the purpose of the changes depicted in the provided KAFE diff.}
\end{enumerate}

As mentioned, the tasks were chosen mainly to ensure the participants explored the diagram's primary features.
In particular, there is no ``correct'' solution to the tasks because we were interested in the way the developers would use the tool and their subjective experiences.
However, we believe the tasks resemble some of the questions a new developer on a project would ask themselves.
By asking the participants to complete tasks \ref{task:projs} and \ref{task:deps}, we guided them towards exploratory analysis, interaction with nodes, and the search feature.
\ref{task:smells} was meant to foster a more narrow \nohyphens{analysis} by suggesting they look at compiler diagnostics and attempt to find any code smells (i.e.,~subjectively bad practices).
We included \ref{task:diff}, the final task, to evaluate an experimental overlay displaying differences between two versions of KAFE.
This overlay is only tangentially relevant to API references and is, thus, outside of the scope of this paper.
It is, however, described in Supplementary Material~\cite{sm}.

\subsection{Participant Observations}

All five developers who participated in the user study made accurate observations about the codebase throughout their 33 to 45 minutes of testing.
In task \ref{task:projs}, they concluded that the most important projects are \texttt{Api} and \texttt{Data}.
Interestingly, they arrived at this conclusion from different angles.
While some argued that it should be the \texttt{Api} project since it is the \textquote[\textbf{E4}]{entry point} to the application, others thought about the task visually and said \textquote[\textbf{E5}]{It would be \texttt{Api}. It takes up most of the pie that has been created.}, referring to the circular diagram as a \textit{``pie''}.
Another participant held a different opinion after examining the \texttt{Data} project:

\begin{displayquote}[\textbf{E3}]
    The \texttt{Kafe.Data} [project] seems like the most important project so far. I was expecting maybe only some [data transfer objects] and data structures, but it appears to contain all the services and the core of the application itself.
\end{displayquote}

\rev{The testers} (\textbf{E1} and \textbf{E3}) also accurately described the \texttt{RUV} project as an integration layer with an external service, and they (\textbf{E1} and \textbf{E5}) \rev{correctly deduced that the} \texttt{Migrator} \rev{project is} a utility for migrating data from a legacy database.

In task \ref{task:deps}, the testers had to enable the \textit{package} entity kind and the \texttt{dependsOn} relation.
They found both standard .NET libraries and external dependencies like FFMpeg and Marten, which led them to accurately assume that the application processes video material and uses the \textit{event sourcing} design pattern.
While looking for code smells (\ref{task:smells}), they located the \texttt{ProjectService} node due to its size and argued that a class of this size is a prime candidate for refactoring.

\subsection{Results}
\label{sec:helveg-results}

\begin{figure}[tb]
  \centering
  \includegraphics[width=0.99\linewidth]{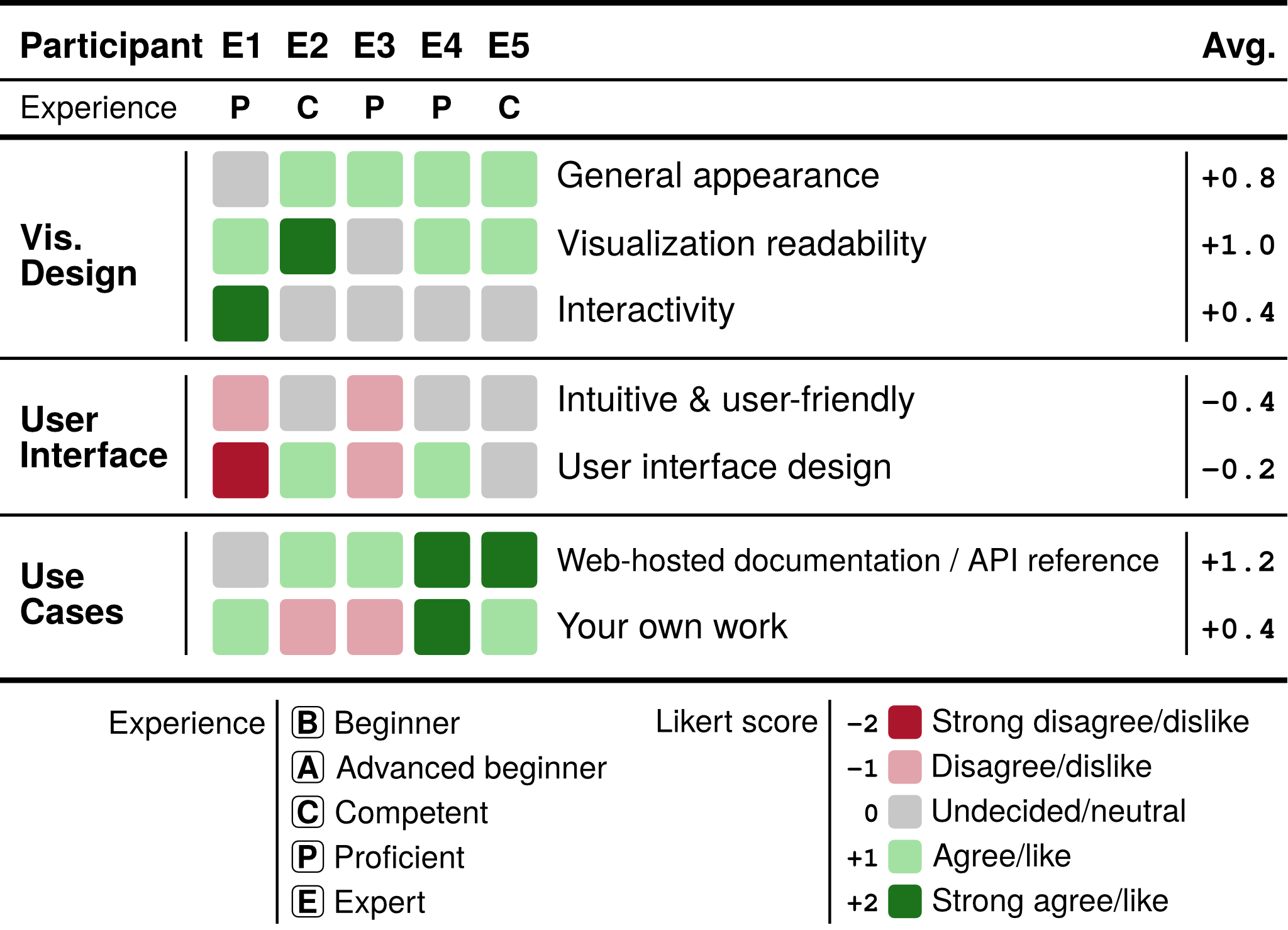}
  \caption{\label{fig:results} Questionnaire results with color-coded Likert scores and an average result for each question.}
\end{figure}

After every concluded testing session, each participant rated their experience with Helveg in several areas on a five-point Likert scale (see results in Figure~\ref{fig:results}).
The participants were mostly satisfied with the visualization design.
They found the glyphs familiar since they already knew the icons from Visual Studio, and they enjoyed the general appearance of the tool as well.
Overall, the tool's visualization was well-received and considered readable and useful, as one of the participants remarked:

\begin{displayquote}[\textbf{E5}]
I enjoyed seeing the graph and the fact that I could imagine what the solution looks like without looking at any piece of code.
\end{displayquote}

One participant thought that the visualization of compiler diagnostics was especially useful.
They remarked that Helveg made it easier to see which projects contained the most errors and warnings compared to the error list found in code editors.
\textquote[\textbf{E4}]{Right here, I was able to see [that] the smell is coming from one concrete place from one project.}, they said.

When asked about comparison with existing tools, the participants responded positively.
One developer considered Helveg to be well-performant, saying that \textquote[\textbf{E2}]{Code Map in Visual Studio usually takes so long}.
Another tester said that \textquote[\textbf{E4}]{For the amount of information that is shown right here, the JetBrains UML diagram would be unreadable for me.}.
A developer (\textbf{E5}) also said they were familiar with the concept of documentation enhanced with a node-link diagram since they already use the Obsidian~\cite{obsidian} note-taking tool.

Besides the positive feedback, the testers expressed dissatisfaction with the intuitiveness of the UI.
They suggested improvements, including tutorials in both written and video form, tooltips placed next to the controls, and hiding some of the options behind an \textit{expert mode}.
Some found the UI's similarity to Visual Studio Code inviting, whereas others commented they felt overwhelmed by \textquote[\textbf{E3}]{the three levels of toolbars}.
The diagram's \textit{interactivity} score was negatively impacted by the overall user experience as well.
One participant indicated that they felt reluctant about interacting with the nodes since \textquote[\textbf{E3}]{there's no point because they're going to be gone upon the next refresh}.
Multiple participants also disliked that the most flexible way to filter the graph was through JavaScript, a language they detested.

While Figure~\ref{fig:results} shows that Helveg is far from perfect, especially regarding user experience, all testers saw its potential use.
Some saw it as a helpful onboarding tool: \textquote[\textbf{E5}]{If someone new started working with me, I would use it to explain the high-level overview and to show parts of the architecture and how they fit together.}.
Others said that \textquote[\textbf{E4}]{It could replace my visualization tooling}.
Even the most critical of testers approved of the tool's basic concept: \textquote[\textbf{E3}]{I like the idea itself. I like it for the API reference especially.}.

\section{Glyph Evaluation}
\label{sec:glyph-evaluation}


Apart from the user study of the prototype, we also evaluated our glyph design (see Section~\ref{sec:glyphs}) to test its readability.
There were six programmers whose C\# programming experience ranged between 1 to 13 years, with a median of 5.5 years.
One of the participants also tested our prototype, as described in the previous section.
The evaluation was a questionnaire with questions about each glyph element: the icons, the outlines, and the effects.
During the evaluation, a researcher was present to clarify any ambiguities and to take note of the participants' behavior.
The questionnaire and its results are attached to Supplementary Material~\cite{sm}.

We evaluated our assumption that VS icons would be familiar to C\# programmers through an image depicting 18 example glyphs with most of the relevant icons.
The questionnaire asked the participants to assign each glyph to a description such as ``public class'' or ``method parameter''.
On average, the participants were able to correctly assign 68.52~\% of glyphs, with a median of 72.22~\%.
This questionnaire section took the most time, from 5 to 16 minutes.

Then, the form explained the meaning of our glyphs' outlines.
Participants were asked to compare several type and type member glyphs, focusing on the presence and width of the outlines.
\rev{These glyphs were shown in a single image to allow the participants to compare them.}
The questions asked about glyphs that, for example, ``represented a type with the most static members'' or the total number of static C\# entities.
The average success rate was 61.11~\%, with a median of 58.33~\%.

The next two sections were dedicated to glyph effects representing compiler diagnostics.
The respondents were first asked to give their opinions on them when the glyphs were isolated and later in a large group, simulating the view in our prototype.
The answers were given on a five-point Likert scale ranging from \textit{Strongly agree} (+2) to \textit{Strongly disagree} (-2).
All participants agreed that the animated effects were a faithful and intuitive representation of compiler diagnostics, with both aspects receiving a score of 1.5.
The question of whether a static effect would be preferable received a score of -0.67.
Only one programmer responded that they would prefer a non-animated effect instead, clarifying that merely different colors of the glyphs would be enough for them.
The results also show that all participants noticed the presence of the fiery effect in a large group (+2) and, likewise, the absence of the smoke (-1.5).
None would consider the depicted codebase to be healthy (-2).

At the end, the questionnaire offered an optional open feedback text field.
Two participants commented on the representation of warnings, saying that \textquote{In [a] larger project, it's hard to distinguish compile errors and warnings.}
One respondent went into detail on their experience, citing that they were \textquote{not used to look at the icons representing types in VS}.
However, they also said the \textquote{visual representation is nice}, although they would need more time and examples to get used to it.


\section{Discussion}

The interviews with C\# programmers and the glyph evaluation show that our diagram can serve as a feasible alternative to an API reference.
However, they also evidence that our approach, and especially our prototype, has several issues.
In this section, we discuss the lessons learned from the evaluation and the work that can be done to address them in the future.

\subsection{Lessons Learned}

\vspace{0.5em}\noindent\fbox{\begin{minipage}{0.98\linewidth}
\vspace{0.1em}
    Our \rev{glyphs and diagram-based navigation} can be used to gain knowledge about a codebase that a typical API reference can hardly provide.

\vspace{0.1em}
\end{minipage}}\vspace{0.5em}

The testers' observations show what can be learned from the diagram.
For example, their discovery of the need to refactor the oversized \texttt{ProjectService} class could be made with a typical API reference only with great difficulty.
In the API reference, the number of type members is not discernible unless one reads every type's page.
Another example is the visualization of compiler diagnostics since, to the best of our knowledge, no API reference generator includes those in its output.
In order to see those, the reader would need to download the source code, which may be unavailable, and compile it themselves, which could require a significant amount of time.

\vspace{1em}\noindent\fbox{\begin{minipage}{0.98\linewidth}
\vspace{0.1em}
    A familiar interface and a set of icons do not necessarily make a UI intuitive.
    
\vspace{0.1em}
\end{minipage}}\vspace{0.5em}

Helveg's evaluation shows a deficiency in its user experience.
This result is not unexpected, considering that the user study was the prototype's first contact with real users.
It shows that although we designed the UI to feel familiar, using recognizable UI concepts and icons, this familiarity alone does not lead to an intuitive UI.
Without any introduction, the tool can seem overwhelming due to its many options.
These options are useful and make Helveg flexible, but they also make the tool more difficult to master.
This difficulty could be managed by equipping Helveg with onboarding features.
The UI already contains a guide panel offering a non-intrusive, optional tutorial.
However, none of the testers opened the panel prior to a hint from the interviewer.
Thus, we conclude that either the UI should be redesigned to be more intuitive or a more prominent tutorial should be added.

The glyph evaluation indicates that while C\# programmers are familiar with the VS icons, they struggle to assign meaning to them.
After the evaluation, a discussion with one of the respondents also revealed that the icons are not used consistently within VS.
Some VS icons can even be confusing since their visual similarity may suggest a connection between them, despite there being none, as is the case with the icon for \textit{fields} and \textit{method parameters}.
Consequently, we believe it to be a good practice to exploit the familiarity of the VS icons but to use them only as a basis for custom visual elements.

\vspace{1em}\noindent\fbox{\begin{minipage}{0.98\linewidth}
\vspace{0.1em}
    A filtering feature must balance its flexibility with its ergonomics.

\vspace{0.1em}
\end{minipage}}\vspace{0.5em}

We also learned that simple features are not always easy to use.
The JavaScript filter mode, which we implemented in Helveg, is simple in its implementation and flexible in its use cases.
However, C\# developers were reluctant and felt uncomfortable writing code in JS.
This shortcoming could be remedied in several ways.
The addition of another search mode that would contain predefined filters is one of them.
With this option, the diagram could be equipped with the most common queries by default.
These queries would additionally serve as samples, making it easier for users to write their own queries.
The JS filter could also be replaced with a \textit{filter builder} interface, allowing users to construct filters more ergonomically at the expense of some loss of flexibility and implementation complexity of this feature.

\vspace{1em}\noindent\fbox{\begin{minipage}{0.98\linewidth}
\vspace{0.1em}
    The prototype neither scales well nor performs well when applied to large codebases.

\vspace{0.1em}
\end{minipage}}\vspace{0.5em}

Although, according to the testers, Helveg performs better than its most well-known alternatives, it does not scale well on projects much larger than KAFE.
KAFE comprises eight projects, resulting in a diagram with at most 3760 nodes when completely unfiltered and expanded.
However, C\# solutions can grow to hundreds of projects and tens of thousands of nodes.
We knew about this scalability issue when we designed the visualization (see Section~\ref{sec:scalability}), so we included flexible filtering features, allowing the user to choose what to focus on.
Without the filtering features, however, the prototype currently cannot handle large solutions.

In the future, we may consider alternatives to the node-link diagram, such as a circular treemap or a hybrid of the two techniques designed for large codebase exploration.
Helveg's performance bottleneck can likely be solved by rendering the visualization using WebGPU~\cite{webgpu}, a future web standard, currently a working draft, that may be able to visualize a much greater number of nodes.

\vspace{1em}\noindent\fbox{\begin{minipage}{0.98\linewidth}
\vspace{0.1em}
    Animated glyph effects accurately represent compiler diagnostics.
    However, the multiple outlines may be too complex.

\vspace{0.1em}
\end{minipage}}\vspace{0.5em}

The glyph effects received an overwhelmingly positive response in our survey (see Section~\ref{sec:glyph-evaluation}).
Although the survey's sample size was small, this result indicates that the effects are suitable visual representations of compiler diagnostics.
However, we believe that glyph outlines may be too complex due to the low success rate of survey questions aimed at comparing them.
While the participants suggested they would get used to them in time, we may replace them with simpler visual elements in the future.

\subsection{Future Work}

There are many ways to extend our approach besides those implied by the lessons learned.
The most obvious extension would be to support another programming language.
Another potential extension can be found in the participants' enthusiastic reaction to the experimental diffing overlay (see Supplementary Material~\cite{sm}).

The data miner can be extended as well.
For example, it could analyze method bodies, leading to a diagram depicting the method call tree, or provide more details about package references, such as whether they are direct or transitive.


\section{Conclusion}

We introduced a visual alternative to a typical API reference, where navigation is handled by a central node-link diagram.
It provides a high-level view with interactive features, allowing in-depth exploratory analysis of a codebase.
The diagram's glyph-nodes build on iconography that most users already know.
We created Helveg, a prototype implementation of this approach.
Through the evaluation with professional software developers, we conclude that Helveg is functional and can provide insight into a codebase, which would be hard to achieve using a typical API reference.
Among possible future directions are improvements to the diagram's scalability, the visual design of its glyphs, and the prototype's user interface and range of supported programming languages.


\section*{Acknowledgements}

We thank all participants of all our surveys and user studies.
We also thank Marco Schäfer for kindly letting us modify his questionnaire based on a SurveyAMC~\cite{surveyamc} template.


\bibliographystyle{IEEEtran}
\bibliography{IEEEabrv,paper}

\end{document}